# Supramolecular Spin Valves

M. Urdampilleta,[1] S. Klyatskaya,[2] J.-P. Cleuziou,[1] M. Ruben,[2,3*] W. Wernsdorfer[1,*]

[1] Institut Néel, associé á l'Université Joseph Fourier, CNRS, BP 166, 38042 Grenoble Cedex 9, France.

[2] Institute of Nanotechnology (INT), Karlsruhe Institute of Technology (KIT), 76344 Eggenstein-Leopoldshafen, Germany.

[3] Institute de Physique et Chimie de Matériaux de Strasbourg (IPCMS), CNRS-Université de Strasbourg, 67034 Strasbourg, France.

**Magnetic molecules possess a high potential as building blocks for the design of spintronic devices.[i,ii] Moreover, the use of molecular materials opens the way for the controlled use of bottom-up, e.g. supramolecular, processing techniques combining massively parallel self-fabrication with conventional top-down nanostructuring techniques.[iii] The development of solid state spintronic devices based on the giant magnetoresistance (GMR),[iv] tunnel magnetoresistance (TMR),[v] and spin valve effects[vi] has revolutionized the field of magnetic memory applications. Recently, organic semiconductors were inserted into nanometer sized tunnel junctions allowing enhancement of spin reversal,[vii,viii] giant magneto-resistance behaviour was observed in single non-magnetic molecules coupled to magnetic electrodes[ix], and the use of the quantum tunnelling properties of single-molecule magnets (SMMs) in hybrid devices was proposed.[ii] Herein, we present an original device in which a non-magnetic molecular quantum dot, made of a single-wall carbon nanotube (SWCNT) contacted with non-magnetic electrodes, is laterally coupled via supramolecular interactions to a $TbPc_2$-SMM (Pc = phthalocyanine), which provides a localized magnetic moment. The conductance through the SWCNT is modulated by sweeping the magnetic field, exhibiting magnetoresistance ratios up to 300% between fully polarized and non-polarized SMMs below 1 K. We thus demonstrate the functionality of a supramolecular spin valve without magnetic leads. Our results open up prospects of circuit-integration and implementation of new device capabilities.**

A standard GMR spin valve is an electronic device, in which two conducting magnetic layers are separated by a non-magnetic layer. Each magnetic layer has a majority spin band, which can propagate into the other layer only if both magnetizations are parallel. The conductance of the device is modulated by the altering magnetic configuration (parallel or antiparallel). Because the layers have different magnetic coercivities, in an increasing external magnetic field the magnetization of one layer switches at different field values than the other, i.e. going from parallel to antiparallel, and finally back to parallel alignment,[iv] thus switching the conductance of the device. The resulting magnetoresistance ratio is defined by MR = ($G_P$ − $G_{AP}$)/$G_{AP}$, where $G_P$ and $G_{AP}$ are the conductances of the spin valve for parallel (P) and antiparallel (AP) alignment. Typical values of MR for metallic spin valves lies in the ten percent range at room temperature. A tunnel barrier between the two layers leads to one hundred percent higher MR values, for instance, in commercially used spin valves in read heads. The two layers can be also connected via a carbon nanotube, leading to typical MR values of about 3% for permalloy single-wall carbon nanotube (SWCNT) spin valves[x,xi] and 60% for $La_{(1-x)}Sr_xMnO_3$ (LSMO) based electrodes.[xii] In all these devices, the layers are classical magnets.

An alternative experimental set-up proposes to use quantum nanomagnets.[ii] It was suggested that if a so-called single-molecule magnet (SMM) is laterally coupled to a contacted SWCNT, its highly anisotropic magnetic moment can influence the current passing through the SWCNT, and thus permitting the readout of the molecular magnetic state by standard conductance measurements. This idea is experimentally realized here. We have measured high magnetoresistance ratios in this supramolecular SMM-SWCNT geometry, in which single quantum nanomagnets act as both magnetic polarizer and analyzer. Remarkably, the presented experimental results demonstrate the electrical detection of the magnetization switching of a single quantum magnet. Out of 130 studied devices, 25 samples showed signals due to magnetic molecules. Seven devices were studied in detail and exemplary data for one of them are presented below (others are shown in Fig. S1 of the SI).

The rare-earth-based SMMs are among the most promising systems for molecular spintronic applications. In *bis*-phthalocyaninato-Terbium (III) complexes, hereafter $TbPc_2$,[xiii] the total magnetic moment is equal to $J = 6$ and originates from both orbital and spin contributions.[xiv.] In its neutral form the $TbPc_2$ molecule represents a two-spin system: the Terbium (III) ion owns an intrinsic magnetic anisotropy while the organic $S = 1/2$ radical is delocalized over the two phthalocyanine ligands ensuring the magnetic coupling to the environment (Fig. 1a). At low temperatures, quantum magnets of the $TbPc_2$-family are characterized by (i) a large magnetic moment in the ground state, (ii) a high zero-field splitting due to their large magnetic anisotropy, (iii) slow relaxation of the magnetization, and (iv) a strong hyperfine coupling.[xv] Additionally, their magnetic properties show a very large spectrum of quantum physics phenomena,[xvi] e.g. quantum tunnelling of magnetization (QTM) between up and down magnetization polarisations and quantum interference between different tunnelling paths.[xvii]

The investigation of the magnetic characteristics of the TbPc$_2$ SMM crystal, using a micro-SQUID at 0.04 K,[xviii] reveals a hysteresis loop entailing magnetic quantum phenomena (Fig. 1b).[xv] The sharp steps within the magnetic hysteresis loop originate from quantum tunnelling at the respective energy level crossings presented in Fig. 1c. Due to the large magnetic anisotropy of Tb(III) ion, the ground state doublet with quantum numbers $J_z = \pm 6$ is well separated from the excited states by several hundreds of Kelvin.[xix,xx] Moreover, the interaction of each ground state with the four nuclear spin states of the Tb(III) ion nuclear spin ($I = 3/2$) leads to several energy level crossings and therefore quantum tunnelling of magnetization (QTM). An additional magnetization reversal mechanism, occurring at higher field, is given by the so-called direct relaxation process, which can be seen as a non-coherent tunnelling event combined with a phonon emission. Therefore, TbPc$_2$ SMM can switch at different external fields, depending on the mechanism involved in the magnetization reversal (see SI, Fig. S3). Recently, it has been shown, that the prominent magnetic properties of TbPc$_2$ SMMs remain robust when attached via supramolecular π-π interactions to sp$^2$-carbon materials, as carbon nanotubes [xxi] and graphene.[xxii] However, it is important to note that each molecule when deposited on the nanotube has slightly different properties because of differences in the structural relaxation. As a matter of fact some of them would have a huge QTM probability whereas others would mostly undergo direct relaxation process.

In order to increase the non-invasive attachment to SWNTs at very low concentration, the TbPc$_2$-SMMs used herein were modified by introducing one pyrene group and six hexyl groups in one of the both Pc-rings (referred hereafter as TbPc$_2$*, see Fig. 1a). Both the pyrene group and the alkyl chains are known to exhibit attractive van der Waals interactions with sp$^2$-carbon materials and are used as anchoring point on the nanotube. Moreover, the steric hindrance induced by this ligand prevents recrystallization of the SMM on the nanotube. The anchoring point steers the supramolecular grafting of the quantum magnet and brings the substituted Pc-ring in direct contact with the SWCNT wall. The latter and the Pc-ring are conjugated systems, which strongly hybridize through π-π interaction.[xxiii] The supramolecular sample geometry is shown in Fig. 1d.

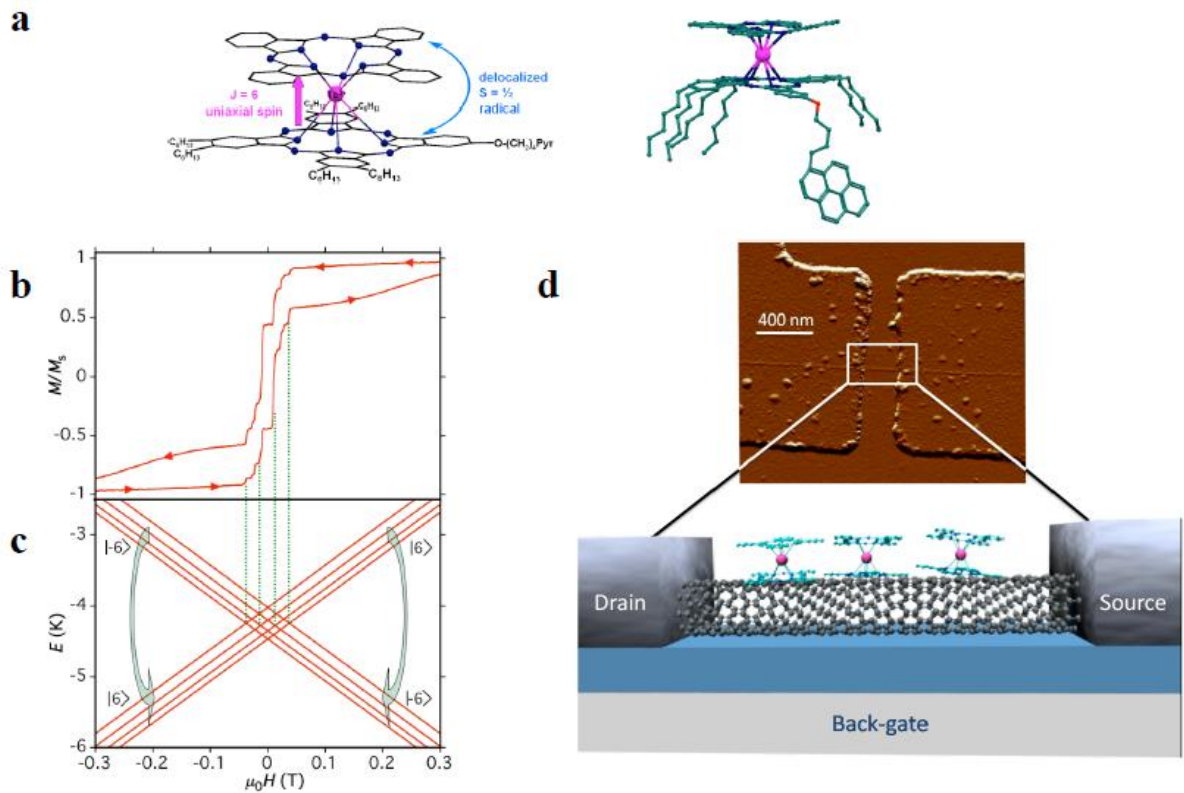

**Figure 1:** Supramolecular spin valve device: **a**, Schematic and molecular representation of the TbPc$_2^*$ quantum nanomagnet. A terbium(III) ion (pink) is coordinated by two phthalocyanine groups; pyrenyl and hexyl substitutions maximize the supramolecular interaction with sp$^2$-carbon materials. The quantum magnet consists of two spin systems: an organic S = 1/2 radical delocalized over the two phthalocyannine rings and a highly anisotropic $J = 6$ spin system localized on the Tb(III) metal ion. **b** Magnetic characteristics of the TbPc$_2^*$-SMM showing the hysteresis loop at 0.04 K for a single crystal of diluted sample measured with a micro-SQUID setup at a field sweep rate of 1 mT/s. The marked steps are induced by quantum tunnelling at energy level crossings presented in **c**. **c**, Zeeman diagram for the lowest states with $J_z = \pm 6$. Those are split by the four nuclear spin states of $I = 3/2$. The calculation was done with the ligand field parameters of Ref. xv. **d**, top, atomic force micrograph of the supramolecular spin valve. The single-wall nanotube sits on a SiO$_2$ surface supported by a back-gate and is connected to palladium source and drain electrodes. **d, bottom,** Scheme of the supramolecular spin-valve architecture (hexyl and pyrenyl groups are omitted by reasons of clarity).

The external magnetic field is applied in the plane of the sample (0° corresponds to the normal to the nanotube axis) and a back gate fine tunes the chemical potential of the SWCNT. The electronic transport characteristics of the supramolecular TbPc$_2^*$-SWCNT set-up are presented in Fig. 2. Conductance maps of d$I$/d$V$ as a function of source-drain voltage $V_{sd}$ and back gate voltage $V_g$ were taken at 0 T (Fig. 2a) and 1 T (Fig. 2b), corresponding to a random and polarized orientations of the TbPc$_2^*$ magnetic moments, respectively. At both magnetic fields, the conductance maps exhibit the features of Coulomb diamonds,[xxiv] typically observed for SWCNT quantum dot in the weak coupling regime.[xxv] Intriguingly, the degeneracy points of most of the Coulomb diamonds are open in an absence of an external magnetic field (0 T, Fig. 2a) and closed in presence of it (1 T, Fig. 2b). This is a clear evidence of the modulation of an extra tunnelling barrier inside the quantum dot. Further insight can be gained when measuring the magnetic field hysteresis loops of the conductance at $V_{sd} = 0$. In Fig. 2c, the magnetic field is swept between 1 T and -1 T resulting in an abrupt switch of the conductance between a relatively high value (~ 1 µS) and a lower one (~ 100 nS). The characteristics of the hysteresis loops are consistent for each of the Coulomb peaks, as shown on Fig. 2d: the difference in conductance value is plotted between trace (from -1 T to 1 T) and retrace (from 1 T to -1 T) for each gate step. The hysteretic behaviour disappears above a bias voltage of about 0.5 mV.

In order to relate the device characteristics with the magnetic signature of the $TbPc_2^*$ quantum magnets, magnetic field studies of the largest conductance jumps were performed. The analysis of the angle dependence of the switching field, plotting the difference between trace and retrace, reveals a very pronounced dependence. The smallest switching fields -- defining the easy axis of magnetization -- were found orthogonal to the SWCNT axis ($x$-direction), whereas along the tube axis ($y$-direction) a hard axis was found (Fig. 3a). Such a very strong angular dependence is in agreement with the Ising-like uniaxial anisotropy of the $TbPc_2$-SMM family. We compare the orientation of the easy axis to the SWCNT axis. It turns out that the $TbPc_2^*$ molecule is orientated with the phthalocyanine rings parallel to the SWCNT wall, which is in accordance with the supramolecular design maximizing the π-π interactions between Pc-ring and the nanotube.

Quantum tunnelling of magnetization represents an inherently stochastic process: the magnetization reversal at a tunnel splitting or via a direct relaxation process is expressed by transition probabilities. By cycling the hysteresis loop many times, the related stochastics of the magnetic moment reversal can be elucidated (Fig. 3b). Increasing temperatures lead to a continuous reduction of the area of the hysteresis loop (Fig. 3c). At 40 mK the switching field is found to be around 0.5 T progressively declining and finally vanishing above 600 mK. The blocking temperature can be extrapolated to be around 1 K, which is in good agreement with recent reports on a $TbPc_2$ submonolayer.[xxvi] Other samples showed similar temperature dependences (see SI).

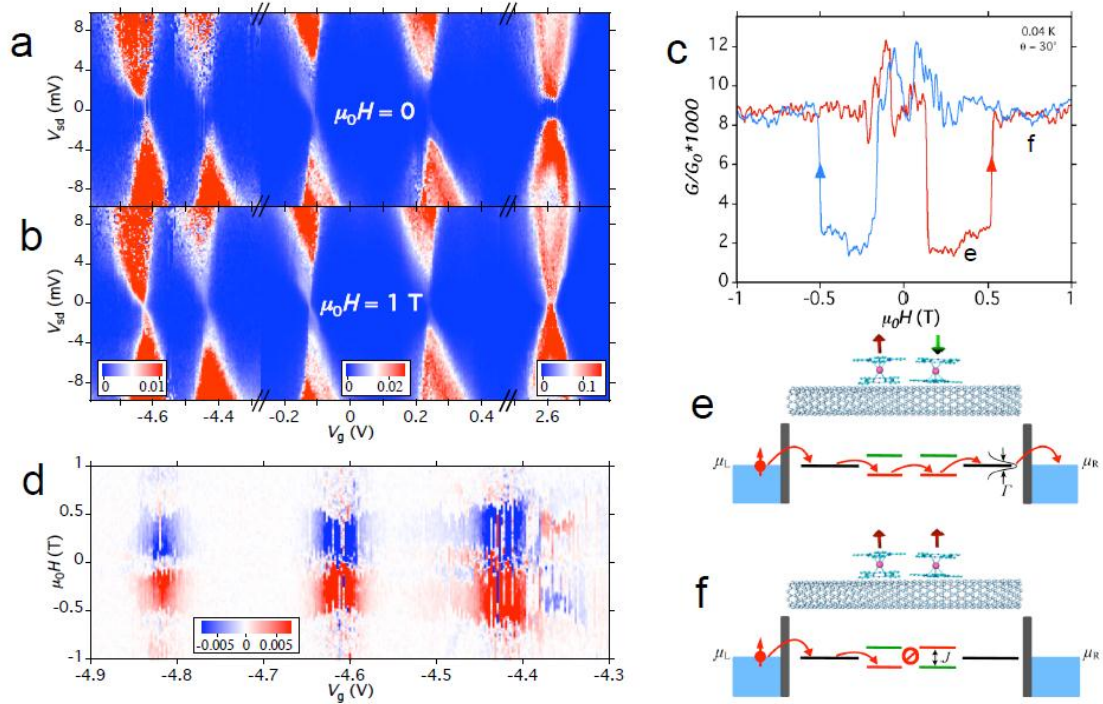

**Figure 2:** Molecular spin valve electronic transport characteristics. **a**, Coulomb map of the differential conductions as a function of source-drain voltage $V_{sd}$ and backgate voltage $V_g$ without magnetic field. $G_0$ refers to the conductance quantum. A gap $\Delta$ is induced at the degeneracy point when all the magnetic moments are randomly oriented. Magnetic molecules induce effective tunnelling barriers in the SWCNT, hindering the electron flow, except in the case where the electrons have enough energy ($V_{sd} > \Delta/2$) to overcome these barriers. **b**, Coulomb map under a magnetic field of 1 T. The gap is closed, and the standard degeneracy points are recovered. It corresponds to the situation where all molecules are polarized approximately in the same direction. **c**, Zero-bias conductance measured as a function of the magnetic field. The red curve corresponds to the conductance under increasing field (-1T to +1T) and the blue curve under decreasing field. The conductance jumps around zero-field are attributed to molecules experiencing quantum tunnelling due to tunnel splittings. The last jumps around ±500 mT are attributed to a direct relaxation process of a single molecule. When all the magnetic momenta are parallel, only one spin carrier can flow easily, whereas when the magnetic moment are antiparallel, the electron flow of both spin carriers is hindered. **d**, Intensity of the magnetic hysteresis measured as a function of the gate voltage. The color code corresponds to the difference between magnetic field trace and retrace, blue corresponds to negative value red to positive and white to zero hysteresis. The most intense region is directly correlated to the Coulomb-peaks. **e-f**, Scheme of the mechanism involving two $TbPc_2^*$-SMM-molecules (A and B) grafted on a SWCNT. With increasing the magnetic field, molecule **A** switches first (**f**), thus leading to an anti-parallel configuration of the spin valve with lowest conductance. Each molecule induces localized states in the nanotube through exchange interaction. The value of this interaction is estimated around J = 0.5 meV. The mismatch between spin levels induces effective tunnel barriers in the SWCNT for both spin polarizations. As a result, the electron flow through the SWCNT is hindered. When molecule **B** switches (**e**), a parallel configuration is recovered and because of the level broadening (~0.2 meV), this configuration leads to high conductance. The hexyl and pyrenyl groups of the $TbPc_2^*$-SMM are omitted by reasons of clarity.

The observation of electronic coupling of the terbium based $J = 6$ magnetic moment with the conduction electrons is explained by the intermediating presence of the S = ½ π-radical delocalized on the organic phthalocyanine ring systems. Thanks to π-π interaction, this radical is supposed to be in close electronic contact with surfaces, e.g. it was shown to lead to Kondo features in scanning probe spectroscopy studies.[xxvii] This means that the wave function of this unpaired spin can easily hybridize with the π-electron density of the nanotube. Moreover, transition metal-Pc systems adsorbed on sp²-graphite surfaces show a pinning of the lowest unoccupied molecular orbital (LUMO) level close to the Fermi-level[xxviii] and recent studies have proven a weak antiferromagnetic exchange coupling of the Tb magnetic moment of TbPc2 into a ferromagnetic Ni-substrate.[xxix] Additionally, the radical ligand state is close in energy to the Tb-4f states,[xxx] enabling an efficient coupling of the terbium magnetic $J = 6$ state with the conduction electron of the SWCNT.

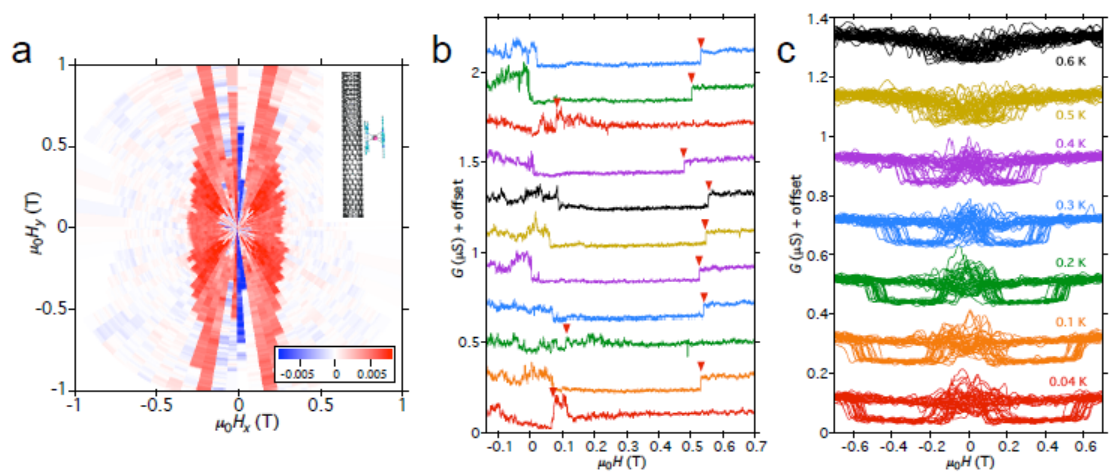

**Figure 3:** Molecular spin valve switching field characteristics. **a,** Angular dependence of the switching field corresponding to the direct relaxation process. The difference between trace (from -1 T to +1 T) and retrace (from +1 T to -1 T) is plotted as a function of the angle of the applied field. The white code corresponds to no difference between trace and retrace for the corresponding field value whereas a red code correspond to a bistable region. This means that in the white region both molecules are polarized in the same way and in the red region the antiparallel configuration is adopted. As a result, the border between the white and the red region corresponds to the switching field of the molecule experiencing a direct relaxation process. It is important to note that the switching field along the *y*-axis cannot be resolved because our magnets are limited to 1 T. This ellipsoidal behaviour has been measured repeatedly on several samples (see SI), and is in agreement with the

Ising like uniaxial anisotropy of the TbPc$_2$*. The x axis can be attributed to the easy axis and the y axis to the hard axis. By comparing the orientation of the easy axis with respect to the nanotube direction it can be deduced that the TbPc$_2$* molecule is flat-landing on the nanotube, as shown in insert. **b**, The conductance is measured 11 times as a function of the magnetic field applied at an angle of 30° with a sweep rate of 2 mT/s. The switching field of the studied molecule is marked by an arrow. Some of the curves present a transition close to zero instead of 0.5 T. This is attributed to a tunnel mechanism at an avoided level crossing. In this case all the TbPc$_2$* molecules have experienced a tunnel transition and the spin valve behaviour disappears (see SI). Most of the tunnel transitions happen at small positive field values, establishing that the Tb nuclear spins are cold enough to involve mainly the lowest magnetic spin states. **c**, 20 hysteresis loops at several temperatures between 0.04 and 0.6 K, using a field sweep rate of 70 mT/s. Few molecules switch around zero field via a tunnel process whereas one molecule switches at higher fields via a direct relaxation process. The blocking temperature is extrapolated to be around 1 K.

The observed magnetoresistance behaviour is explained by an effective tunnelling barrier induced by the magnetic configuration of very few SMM coupled to the SWCNT quantum dot. The average number of molecules was previously determined to be about 4 molecules for a source–drain SWCNT segment of 300 nm.[xxii] The conduction electrons of the SWCNTs are locally influenced through the π-radical mediated exchange mechanism (*vide supra*) of the confined TbPc$_2$*-magnetic moment. For reasons of simplicity, the mechanism will be explained for the most ideal case involving two SMM-molecules (A and B, Fig. 2e-g) and one spin-degenerate conduction channel: each molecule locally lifts the spin degeneracy, and as a result there is a energy mismatch in the antiparallel configuration between the orbital of the same spin ,which a creates an effective tunnelling barrier. In a spin valve picture, the first of the molecules (A) plays the role of spin-polarizer, while the second one (B) acts as spin-analyzer. The presence of two reversal mechanisms (tunnel splitting *versus* direct relaxation process) expressing significantly different tunnelling probabilities (See SI) leads to different switching fields for quantum magnets A and B. In the example presented herein the molecule (A) has a close to 1 QTM probability around zero field (but spread due to the hyperfine coupling) and consequently exhibits magnetic moment flips close to zero field (Fig. 2f). With increasing magnetic field, the second molecule (B) remains still in the opposite magnetic state, rendering an anti-parallel configuration of the spin valve, until this second molecule experiences a direct transition (Fig. 2g), reprogramming the valve into its parallel configuration. Moreover, the quantum behaviour of the second molecule (B) has been studied through its magnetic field sweep rates dependence (Fig. 3c). With decreasing fields, around a rate of 2 mT/s, the spin-valve characteristic disappears in one third of the cases, which is in accordance with the Landau-Zener theory. When molecules A and B tunnel close to zero-field, the spin-valve behaviour vanishes. Importantly, the supramolecular spin valve described here offers two distinct working regimes: A (i) bi-stable and a (ii) sweeping–rate independent regime which allows for functional fine-tuning as well as electronic detection and manipulation of a single magnetic moments. It has to be mentioned that other mechanisms invoking any shift of the Coulomb diamond to explain conductance changes (e.g. magnetocoulomb effect) are refuted by Fig. 2d, where no change of the magnetoconductance sign is turned out.

In conclusion, the working principles of a supramolecular spin valve consisting of quantum magnets (SMMs) and SWCNT components were shown. On reversing an external magnetic field, the device set-up shows magnetoresistance ratios up to 300 % between fully polarized and non-polarized SMMs. Analysis of the switching field angle and temperature dependence has revealed the finger-print like involvement of the magnetic molecules exhibiting Ising-like uniaxial anisotropy and quantum tunnelling phenomena in the conductance of the SWCNT. By demonstrating the functionality of a supramolecular spin valve, our results open an avenue towards the design of operable molecular spintronic devices projecting the implementation of new electrical functionalities, high integration depth with drastically reduced device foot-prints, and additionally, open access to alternative self-fabrication schemes beyond cost-intensive lithographic production.

**Methods**

**Sample Preparation.** SWCNT of a diameter ca. ~1.2 nm grown by laser ablation method at the Rice University were dispersed in a 1-2 mL dichloroethane and sonicated for 1 h. A droplet of this suspension was deposited onto a degenerated p-type silicon wafer with a ~450 nm surface oxide. SWCNTs were located by AFM with respect to predefined markers and then contacted with 50 nm thick Pd by standard electronic lithography with a contact spacing of ~300 nm. The TbPc$_2^*$-SMMs were synthesized as previously reported.[xxi] The supramolecular grafting was carried out by drop casting a TbPc$_2^*$ solution in dichloromethane (DCM) with molarity M=10$^{-8}$ mol.L$^{-1}$ onto the sample. After 5 s, the sample was rinsed in DCM and dried under nitrogen flow. Residual DCM was removed by a second rinse with isopropanol.

**Conductance experiments.** Samples with high resistance (>100 kΩ) at room temperature were selected. The conductance measurements were carried out in a $^3$He/$^4$He dilution refrigerator with a base temperature of 30 mK. Identical stable spin-valve characteristics have been seen on 2 different samples using this method and on 5 samples with the same method but with nanotubes grown by chemical vapour deposition (see SI). The magnetic field in the sample plane was provided by two magnets, generating up to 1 T and up to 0.7 T, respectively. Electrical measurements of interest were performed using a Stanford Research Systems SR-830-DSP lock-in amplifier or an ADWIN source meter in lock-in mode.

**Acknowledgements** This work is partially supported by the DFG-SPP program 1459, ANR-PNANO project MolNanoSpin No. ANR-08-NANO-002, ERC Advanced Grant MolNanoSpin No. 226558, and STEP MolSpinQIP. Samples were fabricated in the NANOFAB facility of the Néel Institute. We thank M. Affronte, F. Balestro, N. Bendiab, L. Bogani, E. Bonet, V. Bouchiat, A. Candini, T. Crozes, E. Eyraud, D. Feinberg, R. Haettel, C. Hoarau, J. Jarvinen, D. Lepoittevin, L. Marty, N-V. Nguyen, T. Novotny, R. Piquerel, V. Reita, C. Thirion, and R. Vincent.

**References**

[i] Rocha, A. R., Garcia-Suarez, V. M., Bailey, S. W., Lambert, C. J., Ferrer, J. & Sanvito, S.: Towards molecular spintronics. Nat. Mat. 4, 335-339 (2005).

[ii] Bogani L. & Wernsdorfer W. Molecular spintronics using single-molecule magnets. *Nature Mat.* **7**, 179 - 186 (2008)

[iii] Lehn J.-M. Supramolecular chemistry: From molecular information toward self-organization and complex matter. *Rep Prog Phys* **67**, 249–265 (2004).

[iv] Baibich, M. N. et al. Giant magnetoresistance of (001)Fe/(001)Cr magnetic superlattices. *Phys. Rev. Lett.* **61**, 2472–2475 (1988)

[v] Binasch, G., Grünberg, P., Saurenbach, F., Zinn, W. "Enhanced magnetoresistance in Fe-Cr layered structures with antiferromagnetic interlayer exchange". *Physical Review B* **39**, 4282 (1989).

[vi] Dieny, B. Giant magnetoresistive in soft ferromagnetic multilayers. *Phys. Rev. B* **43**, 1297–1300 (1991)

[vii] Awschalom, D. D. & Flatté, M. M.: Challenges for semiconductor spintronics. *Nature Phys.* **3**. 153-159 (2007).

[viii] Dediu, V., Hueso, L., Bergenti, I. & Taliani, C. Spin routes in organicsemiconductors. *Nature Mater.* **8**, 707-716 (2009).

[ix] Brede, J., Atodiresei, N., Kuck, S., Lazic, P., Caciuc, V., Morikawa, Y., Hoffmann, G., Blügel, S., Wiesendanger, R. Spin- and Energy-Dependent Tunnelling through a Single Molecule with Intramolecular Spatial Resolution. Phys. Rev. Lett. 105, 047204 (2010).

[x] Sahoo, S., Kontos, T., Furer, J., Hoffmann, C., Graber, M., Cottet, A., & Schonenberger, C. Electric field control of spin transport. *Nature Phys.* **1**, 99-102 (2005).

[xi] Aurich H., Baumgartner A., Freitag F., Eichler A., Trbovic J. & Schoenenberger C.; Permalloy-based carbon nanotube spin-valve. *Appl. Phys. Lett.* **97**, 153116 (2010).

[xii] Luis E. Hueso, et al. Transformation of spin information into large electrical signals using carbon nanotubes Nature **445**, 410-413(2007)

[xiii] Ishikawa, N., Sugita, M., Tanaka, N., Ishikawa, T., Koshihara, S. Y. & Kaizu, Y. Upward temperature shift of the intrinsic phase lag of the magnetization of Bis(phthalocyaninato)terbium by ligand oxidation creating an S = 1/2 spin. *Inorg. Chem.* **43**, 5498 (2004).

[xiv] Stepanow, S. et. al. Spin and orbital magnetic moment anisotropies of monodispersed bis(phthalocyaninato) terbium on a copper surface, *J. Am. Chem. Soc.* **132**, 11900–11901 (2010).

[xv] Ishikawa, N., Sugita, M. & Wernsdorfer, W.: Quantum tunnelling of magnetization in lanthanide single-molecule magnets: bis(phthalocyaninato)terbium and bis(phthalocyaninato)dysprosium Anions. *Angew. Chem. Int. Ed.* **44**, 2931-2935 (2005)

[xvi] Christou, G.; Gatteschi, D.; Hendrickson, D. N.; Sessoli, R. Single-molecule magnets. *MRS Bull*. **25**, 66-71 (2000).

[xvii] Wernsdorfer, W. & Sessoli, R.: Quantum phase interference and parity effects in magnetic molecular clusters. *Science* **284**, 133-135 (1999).

[xviii] Wernsdorfer, W. From micro- to nano-SQUIDs: applications to nanomagnetism. *Supercond. Sci. Technol.*, **22**, 064013 (2009).

[xix] Ishikawa, N.; Sugita, M.; Okubo, T.; Tanaka, N.; Iino, T.; Kaizu, Y. Determination of Ligand-Field Parameters and f Electronic Structures of Double-Decker Bis(phthalocyaninato)lanthanide Complexes *Inorg. Chem.* **42,** 2440-2446 (2003).

[xx] Zopellaro, G. et. al. Spin Dynamics in the negatively charged Terbium (III) bis phthalocyaninato Complex, *J. Am. Chem Soc*. **131**, 4387-4396 (2009).

[xxi] Klyatskaya, S., Galan-Mascaros, J. R., Bogani, L., Hennrich, F., Kappes, M., Wernsdorfer, W. & Ruben, M. Anchoring of rare-earth-based single-molecule magnets on single-walled carbon nanotubes *J. Am. Chem. Soc.* **131**, 15143–15151(2009).

[xxii] Lopes M., et al. Surface-Enhanced Raman Signal for Terbium Single-Molecule Magnets Grafted on Graphene. *ACS Nano* **4**, 7531–7537 (2010).

[xxiii] Wang et al. Single-walled carbon nanotube/cobalt phthalocyanine derivative hybrid material. J. Mater. Chem., 21, 3779, (2011).

[xxiv] Hanson, R., Kouwenhoven, L.P., Petta, J.R., Tarucha S. & Vandersypen L.M.K.: Spins in few-electron quantum dots. Rev. Mod. Phys. 79, 1217-1265 (2007).

[xxv] Sapmaz, S.; Jarillo-Herrero, P.; Kouwenhoven, L. P.; van der Zant, H. S. J. Quantum dots in carbon nanotubes. *Semicond. Sci. Technol.* **21** No 11 (2006) S52-S63

[xxvi] Otero, L. M., Caneschi, A. & Sessoli; R. X-Ray Detected Magnetic Hysteresis of Thermally Evaporated Terbium Double-Decker Oriented Films *Adv. Mat.* 5488-5493 (2010).

[xxvii] Katoh, K. et. al. Direct Observation of Lanthanide(III)-Phthalocyanine Molecules on Au(111) by Using Scanning Tunnelling Microscopy and Scanning Tunnelling Spectroscopy and Thin-Film Field-Effect Transistor Properties of Tb(III)- and Dy(III)-Phthalocyanine Molecules *J. Am. Chem. Soc.* **131**, 9967 (2009).

[xxviii] Gopakumar, T. G., Müller, F., Hietschold M. STM and STS studies of planar and non-planar naphthalocyanines on graphite I & II. *J. Phys. Chem B* **110**, 6051, 6060 (2006)

[xxix] Lodi Rizzini, A. et al. Coupling Single Molecule Magnets to Ferromagnetic Substrates. Physical Review Letters 107, 177205 (2011).

[xxx] Vitali, L.; Fabris, S.; Conte, A. M.; Brink, S.; Ruben, M.; Baroni, S.; Kern, K.; Electronic Structure of Surface-supported Bis(phthalocyaninato) terbium(III) Single Molecular Magnets. *Nano Lett* **8**, 3364-3368 (2008).